*Communication*

# Fairness Is an Emergent Self-Organized Property of the Free Market for Labor

**Venkat Venkatasubramanian**

Laboratory for Intelligent Process Systems, School of Chemical Engineering, Purdue University, West Lafayette, IN 47907, USA; E-Mail: venkat@ecn.purdue.edu

**Abstract:** The excessive compensation packages of CEOs of U.S. corporations in recent years have brought to the foreground the issue of fairness in economics. The conventional wisdom is that the free market for labor, which determines the pay packages, cares only about efficiency and not fairness. We present an alternative theory that shows that an ideal free market environment also promotes fairness, as an emergent property resulting from the self-organizing market dynamics. Even though an individual employee may care only about his or her salary and no one else's, the collective actions of *all* the employees, combined with the profit maximizing actions of *all* the companies, in a free market environment under budgetary constraints, lead towards a more fair allocation of wages, guided by Adam Smith's invisible hand of self-organization. By exploring deep connections with statistical thermodynamics, we show that entropy is the appropriate measure of fairness in a free market environment which is maximized at equilibrium to yield the lognormal distribution of salaries as the fairest inequality of pay in an organization under ideal conditions.

## 1. Introduction

The excessive compensation packages of CEOs of U.S. corporations in recent years have brought to the foreground the issue of fairness in economics. The typical response to the question "What is fair pay for a CEO?" is that whatever the free market for labor, driven by the forces of supply and demand, determines it to be. The conventional wisdom, however, is that the free market cares only about efficiency and not fairness. Is this view correct? This is the question we examine in this paper.

The fairness question is a challenging one as the term fairness is often used quite broadly and can mean different notions in different contexts, as can be seen from the extensive literature on this subject (we cite here only a few select papers [1-6] as a representative sample). What is morally fair can be different from what is fair in an economic sense. For instance, fairness based on moral principles would require us to recognize all human beings as equals, but does this imply that everyone should receive equal salary irrespective of their contributions in an organization? Our sense of fairness from an economic perspective would disagree. Our intuitive sense of fairness in wages suggests that one's reward should be commensurate with the value of one's contribution.

Even within the economic context, there are a variety of measures of fairness in use. For instance, given an unequal distribution of salaries among employees in an organization, some commonly used measures are maximin fairness, proportional fairness, Gini coefficient, Theil index and so on. In maximin fairness, the objective is to maximize the salary of the least well off employee in the organization. This is the notion of fairness advocated by Rawls [7]. On the other hand, proportional fairness strikes a balance between maximizing the total productivity of the organization and achieving maximin fairness. The Gini coefficient is generally defined based on the Lorenz curve, as the ratio of the area that lies between the line of equality and the Lorenz curve over the total area under the line of equality. The Theil index is an entropy based measure, which we will address in detail in subsequent sections.

For most fairness measures, the promotion of fairness in a system *requires a central authority* to care about the weakest agent(s) in the system and promote fairness by enforcing a fairness policy utilizing the chosen measure. For example, in the Rawlsian framework, the state, which acts as the central authority, is needed to promote fairness by enforcing the maximin measure based fairness policy. However, in an ideal free market environment there is *no such central authority*. Each economic agent is interested in maximizing only his or her utility or profit and does not care about anyone else's. Does such a free market environment for labor care about fairness in salary distribution? Superficially, as the conventional wisdom holds, the answer seems to be in the negative. However, we show that a deeper analysis reveals that an ideal free market for labor does indeed promote fairness, as an *emergent* property resulting from the self-organizing dynamics of the market environment, which can be quantified to predict the ideal maximally fair distribution of salaries at equilibrium.

## 2. Emergence of Fairness through Self-organizing Free Market Dynamics

Consider a competitive, dynamic, free market environment comprising of a large number of utility maximizing agents as employees and profit maximizing agents as corporations. We assume an ideal environment where the market is perfectly competitive, transaction costs are negligible, there are no externalities, and market participants have perfect information. In this ideal free market, employees are free to switch jobs and move between companies in search of better utilities. Similarly, companies are free to fire and hire employees in order to maximize their profits. We also assume that both utility and profit are strictly functions of money alone and nothing else. In addition, we do not consider the effect of taxes.

We also assume that a company needs to retain all its employees in order to survive in this competitive market environment. Thus, a company will take whatever steps necessary, allowed by its constraints, to retain all its employees. Similarly, all employees need a salary to survive and that they will do whatever that is necessary, allowed by certain norms, to stay employed. We assume that neither the companies nor the employees engage in illegal practices such as fraud, collusion, and so on.

In this ideal free market, consider a company *A* with *N* employees and a salary budget of *M*, with an average salary of $S_{ave} = M/N$. Let us assume that there are *k* categories of employees -- ranging from secretaries to the CEO, contributing in different ways towards the company's overall success and value creation. All employees in category *i* contribute *value* $V_i$, $i \in \{1, 2, \ldots k\}$, such that $V_1 < V_2 < \ldots < V_k$. Let the corresponding value at $S_{ave}$ be $V_{ave}$, occurring at category *s*. Since all employees are contributing unequally, some more some less, they all need to be compensated differently, commensurate with their relative contributions towards the overall value created by the company. Instead, *A* has an egalitarian policy that all employees are equal and therefore pays all of them the same salary, $S_{ave}$, irrespective of their contributions. The salary of the CEO is the same as that of an administrative assistant in the mail room. This salary distribution is a sharp vertical line at $S_{ave}$, as seen in Figure 1(a), the Kronecker delta function of magnitude *N*, given by:

$$f(S) = N\delta_{is}, \text{ where } \delta_{is} = 1, \text{ if } i = s \text{ and } \delta_{is} = 0, \text{ if } i \neq s$$

As noted, while this may seem fair in a social or moral justice sense, clearly it is not in an economic sense. If this were to be the only company in the economic system, or if *A* is completely isolated from other companies in the economic environment, the employees will be forced to continue to work under these conditions as there is no other choice.

However, in an ideal free market system there are other choices. Therefore, all those employees who contribute more than the average–*i.e.*, those in value categories $V_j$ such that $V_j > V_{ave}$ (e.g., senior engineers, vice presidents, CEO) would feel that their contributions are not fairly valued and compensated for by *A*, and will therefore be motivated to leave for other companies where they are offered higher salaries. Hence, in order to survive *A* will be forced to match the salaries offered by others to retain these employees, thereby forcing the distribution to spread to the right of $S_{ave}$, as seen in Figure 1(b).

**Figure 1.** Spreading of the salary distribution under competition.

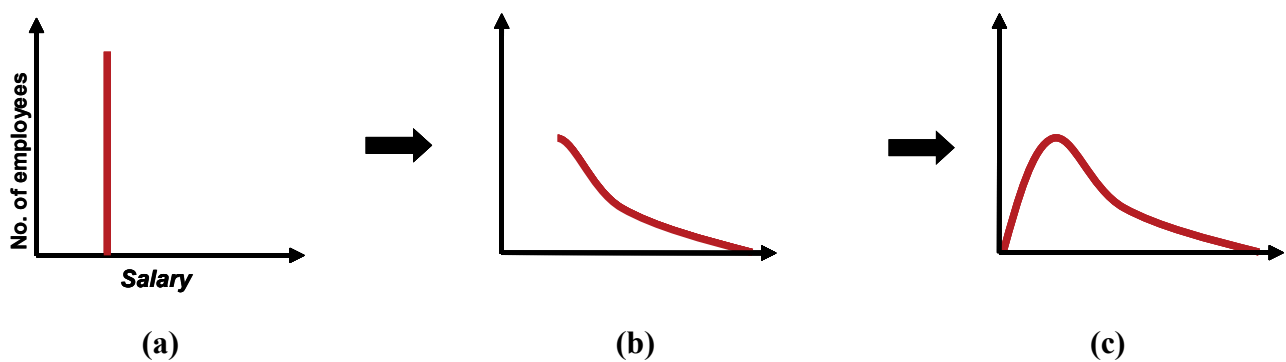

(a)  (b)  (c)

At the same time, the generous compensation paid to all employees in categories $V_j$ such that $V_j < V_{ave}$, will motivate candidates with the relevant skill sets (e.g., low-level administration, sales and marketing staff) from other companies to compete for these higher paying positions in *A*. This competition will eventually drive the compensation down for these overpaid employees forcing the distribution to spread to the left of $S_{ave}$, as seen in Figure 1(c). Eventually, we will have a distribution that is not a delta function, but a broader one where different employees make different salaries depending on the values of their contributions. The funds for the higher salaries now paid to the formerly underpaid employees (*i.e.*, those who satisfy $V_j > V_{ave}$) come out of the savings resulting from the reduced salaries of the formerly overpaid group (*i.e.*, those who satisfy $V_j < V_{ave}$), thereby conserving the total salary budget *M*.

Thus, we see that concerns about fairness in pay causes the emergence of a more equitable salary distribution in a free market environment through its self-organizing, adaptive, evolutionary dynamics and that the spread of the distribution is closely related to fairness in relative compensation. The point of this analysis is not to model the exact details of the free market dynamics but to show that the notion of fairness plays a central role in the emergence and spread of the salary distribution through the free market process. Even though an individual employee cares only about his or her utility (salary) and no one else's, the collective actions of *all* the employees, combined with the profit maximizing survival actions of *all* the companies, in an ideal free market environment of supply and demand for talent, under budgetary constraints, lead towards a more fair allocation of wages, guided by Adam Smith's invisible hand of self-organization.

Observe that even though the initial delta distribution was Pareto efficient in salaries in the Walrasian sense, along the lines of the first and second fundamental theorems of welfare economics (*i.e.*, all the salaries add up to *M*), it did not stay in that state as its final equilibrium state. The utility maximizing behavior of the employees, combined with the profit maximizing behavior of the company, forced the salary distribution to evolve towards its equilibrium state where the salary dispersion is more fair. Thus, we see that Pareto efficiency alone does not guarantee equilibrium. So, what is the condition for attaining equilibrium? This, and the question of what would be an appropriate measure of fairness in an ideal free market system, we address next.

## 3. Entropy as the Appropriate Measure of Fairness in an Ideal Free Market System

As mentioned, there are different measures of fairness such as maximin, proportional, Gini coefficient, Theil index, *etc*. *What then would be the appropriate measure of fairness in an ideal free market environment?* In other words, what measure captures the kind of fairness promoted by the ideal, self-organizing, free market dynamics as discussed in the previous section. Recall that in the ideal free market environment there is *no central authority* to enforce those fairness policies which require the presence of such an authority to compute and enforce the chosen fairness measure. This right away would eliminate measures such as maximin, proportional, Gini coefficient and others which require such a presence.

To identify the fairness measure we seek, we proceed in two directions, with both leading to the same final result. The first is an intuitive approach analogous to the statistical thermodynamic analysis of the evolution of the energy distribution of molecules. The second is an axiomatic approach following the work of Lan *et al*. [8].

*3.1. Entropy as a Measure of Fairness: Statistical Thermodynamics Approach*

In statistical thermodynamics, one analyzes the evolution of many-particle systems by using the concept of *phase space* [9,10]. The phase space is an abstract multi-dimensional space of positions and momenta. For instance, the phase space of *N* mono-atomic gas molecules in a closed container is the *6N* dimensional space specifying the three positions and three momenta for each molecule. A point in this space specifies the positions and momenta of all the molecules, thus representing the state of the system at some time *t*.

Consider an *isolated* thermodynamic system of *N* gas molecules and a total energy of *E*, with an average energy of $E_{ave} = E/N$. In such an isolated system, both *N* and *E* are conserved. Now imagine starting this system at *t=0* in the initial state where all molecules are assigned to have exactly the same energy $E_{ave}$. In time, the system will evolve from this initial state of a Kronecker delta distribution in energy to one where the energy distribution is more spread out as a result of intermolecular collisions which cause varied energy exchanges among the molecules.

Similarly, we can define the *economic* phase space of the company *A* as an *N*-dimensional space with each dimension representing the salary $S_i$ of an employee *i*. In reality, this space will be more complicated with additional dimensions that capture other features such as titles, awards, education, experience, *etc*., but let us restrict ourselves to salary as it is the dominant feature. A point in this ideal phase space represents the state of the company at some time *t* by specifying the salaries of all the employees. Company *A* is an *isolated* system, in the thermodynamic sense, with respect *N* and *M* (*i.e.*, these are conserved) but an *open* system with respect to *information exchanges* regarding salaries and job openings elsewhere in the free market economic environment.

The role of information exchanges is an important distinction between thermodynamic systems and economic systems. In thermodynamics, we have three classes of systems: *isolated* (both total energy *E* and number of molecules *N* are fixed), *closed* (only number of molecules *N* is constant, *E* can vary due to exchanges with the surrounding environment), and *open* (both *N* and *E* vary due to exchanges with the environment). Thus, there are only two kinds of exchanges: energy and matter. However, in our free market environment, we have an additional possibility due to information exchange. We can have a situation where a system is *isolated* with respect to money and employees (*i.e.*, these are fixed as it is in our case) but *open* with respect to information exchanges. Note, however, that information exchange is not subject to conservation laws like those of energy and matter.

As the salaries of the employees increase and decrease, as discussed in Section 2, the company adapts, evolves, and wanders stochastically through the phase space. This process is essentially the same as how *N* molecules of a gas in an isolated system get distributed in their phase space of *6N* dimensions of positions and momenta. Of course, the mechanistic details of the evolution of the two systems are quite different. For molecules, the system evolves through molecular collisions governed by Newton's laws of motion and conservation laws. For employees, it is through the exchange of information between *A* and the free market environment driven by the needs of the rational agents to maximize utility and profit. Information exchanges are the "collisions" in this regard.

We shall now present an intuitive analysis of the dynamics of our ideal free market. To illustrate the key conceptual insights, let us stipulate some numbers for the situation of company *A* discussed in Section 2. Let $N = 1,000$ and $M = \$60$ million. Let us now define the *microstate* and the *macrostate* of this system. Specifying the salaries for each employee, *i.e.*, $\{S_1, S_2, ..., S_{1000}\}$, at any given time $t$ corresponds to defining the *microstate* of *A*. Specifying how many employees are in *each* of the *k salary categories*, at any given time, corresponds to defining the *macrostate* of *A*. Generally, for a large *N*, there will be an extremely large number of macrostates. Most macrostates, in turn, would contain an unimaginably large number of different microstates. This property is known as *multiplicity (W)* in statistical mechanics. The Kronecker delta function configuration presented above is a unique exception, a macrostate that has only one microstate. As the company adapts and evolves over time, it will wander from one microstate to another. This does not mean, however, it will necessarily move from one macrostate to another as many different microstates could correspond to the same macrostate due to multiplicity.

Let us now consider a free market environment consisting of a large number of companies (*A, B, C, etc.*), each one with the same *N* and *M* as *A*. They are all identical replicas of *A*, except that each one is in a *different microstate*. This is the equivalent of the *microcanonical (NVE) ensemble* in *statistical thermodynamics* for a free market environment. As noted, many of the microstates may correspond to the same macrostate. In this ensemble, consider two companies *A* and *B*, shown in Figures 2(a) and 2(b) (the figures are not drawn to scale, as it is not important for our discussion). *A* is in the macrostate of all 1,000 employees making the same salary of \$60,000/year. *B* is in the macrostate of 800 employees making \$30,000/year and the remaining 200 at \$180,000/year. Both satisfy the $N = 1,000$ and $M = \$60$ million constraints. In 2(a) only one category has employees with none in all other categories. This macrostate can be realized in only one way ($W = 1$) and, therefore, this macrostate has only one microstate. In 2(b) two categories are active. This macrostate can be realized in $10^{220}$ ways, through the various permutations of the employees, thus resulting in a multiplicity of $10^{220}$ microstates (we show below how multiplicity is calculated). Let us further assume that there are three classes of employees: low skilled (contributing value $V_{low}$), medium skilled ($V_{med}$), and high skilled ($V_{high}$), such that $V_{low} < V_{med} < V_{high}$, implying that $S_{low} < S_{med} < S_{high}$. Let us further stipulate that the number of low skilled employees is 500, medium skilled 300, and high skilled 200.

**Figure 2.** Free market interaction between A and B: Initial macrostate.

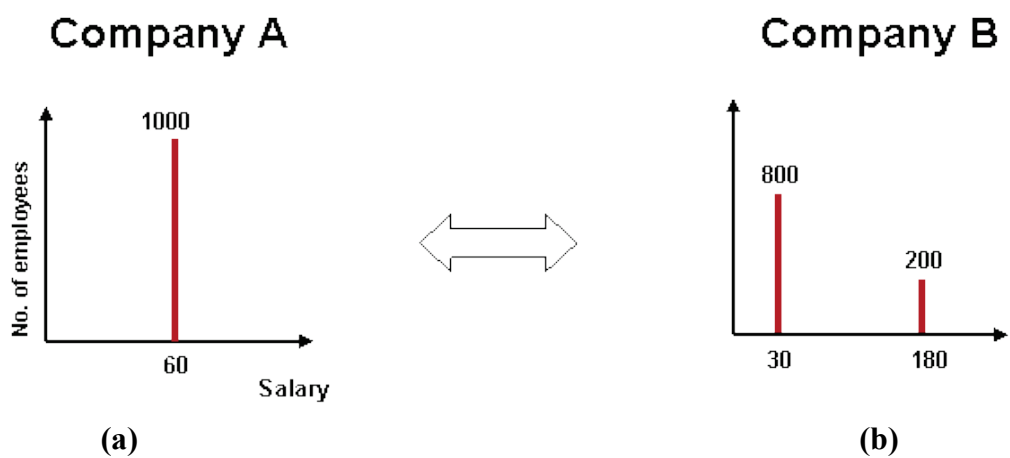

(a)                                                                                           (b)

As these two companies interact, as shown by the double headed arrow, in the free market environment, the 200 high skilled employees in *A* (call it *A-high*), making $60K each, will be motivated to increase their utilities and consider a move to *B* for the higher salary ($180K). Similarly, *B* sees an opportunity to reduce its costs by firing all of its 200 high skilled employees (*B-high*) and hiring the *A-high* group at a lower salary. But in order to attract *A-high B* has to offer a salary that is more than what *A-high* makes currently (*i.e.*, $60K each) as an incentive. However, to *maximize its profit*, *B* will offer *A-high* the *smallest* incentive possible, say, $60K+$\varepsilon$, where $\varepsilon$ is some small amount arbitrarily close to zero (say, $1). Likewise, in order to motivate *B* to hire *A-high* it will offer to work for a salary that is less than what *B-high* makes as an incentive. But, as before, in order to *maximize its utility*, *A-high* will offer the *smallest* incentive possible, and propose a salary of $180K-$\varepsilon$. If both parties don't budge from their respective initial offers no further action takes place and we reach a stalemate.

But let us suppose that they negotiate, as it typically happens in a real free market environment, and agree to split the difference 50:50. Thus, they agree on a salary of $120K/year. *B* decides to fire *B-high* and hire *A-high* at $120K/yr each. *A-high* also decides to leave *A* for *B*. But the *B-high* group do not want to lose their jobs and agree to continue to work at *B* for less salary at $120K/yr. At the same time, to avoid losing the *A-high* group of employees and go out of business, *A* decides to match *B*'s offer and agrees to pay them $120K/yr. Thus, a new category, at $120K, appears in *A* as well as in *B* as seen in Figures 3(a) and 3(b).

In a similar fashion, the 300 medium skilled workers in *B* (*B-med*) making $30K/yr, will try to move to *A*, and, likewise, *A* will try to fire the 300 *A-med* employees and hire *B-med* for a lower salary. As in the previous case, this will lead to the creation of a new category at $45K for 300 employees in both *A* and *B*, as shown in Figures 3(a) and 3(b).

**Figure 3.** Free market interaction between A and B: Evolution of a new macrostate.

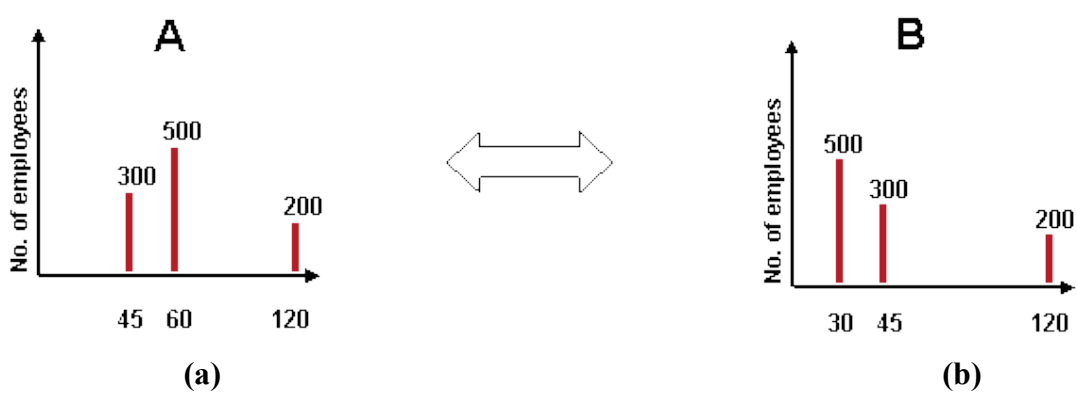

(a)  (b)

Finally, the 500 low skilled employees in *B* (*B-low*) making $30K/yr will try to move to *A* in order to make more. This will result in the addition of another 500 employees to the $45K category and the elimination of the $30K and $60K categories, leading to the final distributions seen in Figures 4(a) and 4(b). These are the final macrostates of *A* and *B* in this context.

Note that these adjustments and alignments do not happen sequentially but simultaneously. Thus, we see that the initial delta function distribution in Figure 2(a) has finally led to the emergence of a bimodal distribution through the free market dynamics. Note that not only *A* changed but *B* also adapted in order to survive. Now, both *A* and *B* are in equilibrium with each other and no further adjustment takes place between them.

**Figure 4.** Free market interaction between A and B: New equilibrium macrostate.

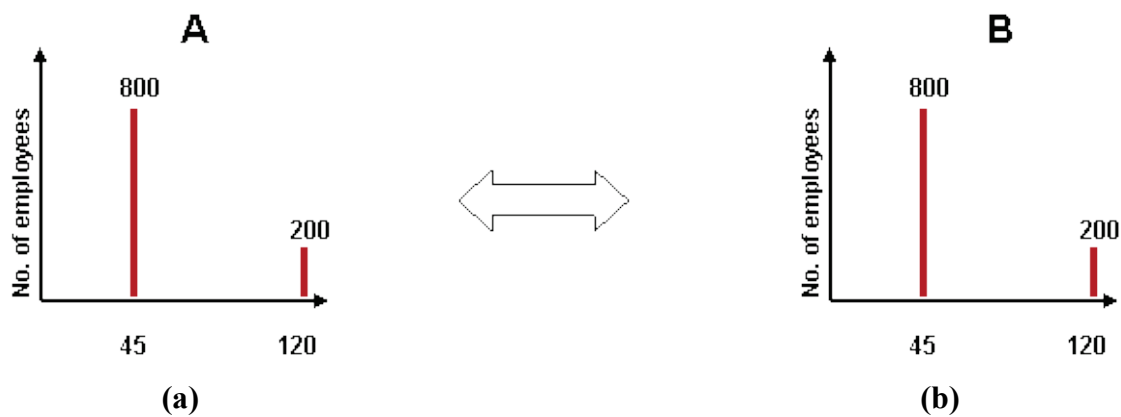

Now, let us examine the situation when this new *A* comes in contact with another company *C* in the ensemble as shown in Figures 5(a) and 5(b). Carrying out the aforementioned analysis, we see that this leads to the final distributions seen in Figures 6(a) and 6(c), which are the new final macrostate.

**Figure 5.** Free market interaction between A and C: Initial macrostate.

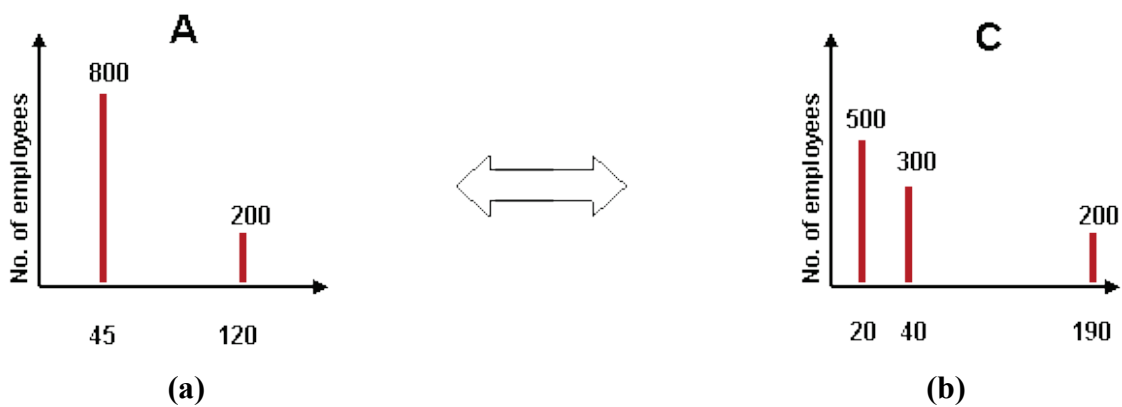

**Figure 6.** Free market interaction between A and C: New equilibrium macrostate.

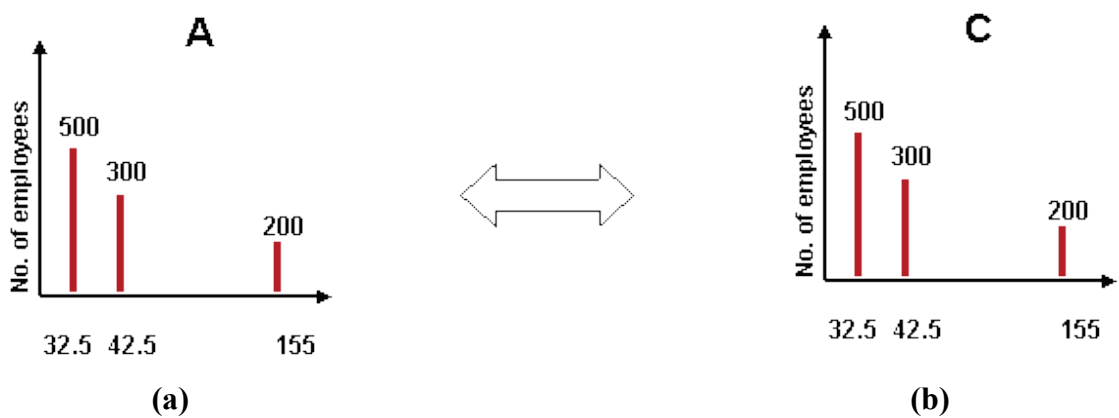

Thus, we see that through such interactions in the free market ensemble companies will evolve, create some new categories, eliminate some other categories, adapt the salaries for the categories, and finally settle in some equilibrium macrostate that naturally emerges as the final resting state. The key insight here is that if a particular value (*i.e.*, salary) category contains employees making varying value contributions (e.g., as in Figure 2(a), where three different types of employees, *A-low*, *A-med,* and *A-high,* are all classified as one salary category), the poorly paid "higher-value" employees will be motivated to seek opportunities elsewhere. This will force all the interacting companies to suitably adapt as we saw in our example above. These self corrections will continue to happen until each category, under the *N* and *M* constraints, finally ends up with a *homogenous* subset of employees (meaning that they are all contributing the same value) all of whom receive the same salary appropriate for that category. Thus, the ideal free market ensures that all employees are treated fairly within their respective value categories (which is the same as salary categories as salaries are proxies of the values contributed by the employees)–*i.e.*, *intra-category fairness* emerges as a natural consequence of the free market dynamics.

But how about *inter-category fairness*? To gain insights about this let us consider another scenario where company *A* starts with a *uniform* distribution of employees over all the possible *k* categories. Let N = *1,000* and *k= 10,* and, therefore, we have 100 employees in each category. We have 100 CEOs (corresponding to the *k = 10* category), 100 vice presidents (*k=9*), 100 engineers (say, *k=5*), *etc*. What this distribution *assumes* is that the total value to be created by *A*, in terms of its products and services, would require such a uniform assignment of employee resources. Obviously, *A* only needs one CEO, perhaps three or four vice presidents, and many more engineers and other personnel. Let us say that the *k = 5* category really needs 200 engineers because of the amount of work needed to be executed at that level, but was assigned only 100.

Since the amount of work to be done, and hence the value created, in the CEO or vice president categories are already specified and limited (as they are in the other categories), one of several work patterns would emerge: all 100 CEOs slack off and contribute 1% each for the total value created by the CEO category, or one CEO creates all the value and the others contribute nothing, or some other combination some contributing and some not. At any rate, not everyone is contributing 100% of the value each, but all of them are getting paid 100% of the CEO salary each.

On the other hand, in *k = 5* category, the 100 engineers work twice as hard to create value that nominally requires the output of 200 engineers. However, they are all paid the salary that is specified for the *k=5* category and not more. While there is *no intra-category unfairness* in the *k = 5* category (since all these employees are contributing the same value and getting paid the same salary), there is *inter-category unfairness* felt by the *k = 5* category engineers. These employees will resent the unfairness of what the CEOs and VPs are getting away with and hence will leave for other companies where there is no such glaring inter-category disparity. Obviously, this can happen at all categories, not just at *k = 5*. As a result, *A* will not survive unless it adapts and reallocates its employee resources more sensibly. We see that when employee resources are not properly allocated across the different categories issues of fairness automatically creep up and the free market tries to correct such imbalances. Since all the companies in the free market will be self correcting in this manner, the distribution will become more and more fair, iteratively, and will eventually reach an equilibrium macrostate where the salary distribution is maximally fair with respect to both intra-category and inter-category under the given constraints.

Thus, we see how the different employee classes, based on their value contributions (which depend on their talents, education, experience, skills, *etc.*), naturally emerge and self organize to reach the equilibrium macrostate, guided by Adam Smith's invisible hands, through the dynamics of the free market environment, such that every employee feels he or she is fairly compensated for the value contribution he or she has made. When all companies in the free market reach such a classification of their employees *equilibrium is reached* as there is no incentive left for any employee (or company) to switch jobs (or employees) as we see in Figures 4 or 6. Thus, the equilibrium distribution is maximally fair. As the companies evolve, adapt, and stochastically search for this equilibrium macrostate that is the optimal compromise under the given constraints, they will settle into a macrostate that can be arrived at in most number of ways–*i.e.*, one with the maximum multiplicity, *W*, subject to the *N* and *M* constraints. This is given by the well known result in statistical mechanics and information theory [9-11,15]:

$$W = \frac{N!}{n_1! n_2! \ldots n_k!}$$

$$Max\ W = Max\ \frac{N!}{n_1! n_2! \ldots n_k!}$$

where $n_i$ is the number of employees in the value category $i$, subject to the constraints:

$$\sum_{i=1}^{k} n_i = N$$

$$\sum_{i=1}^{k} n_i S_i = M$$

Using Stirling's approximation, for large *N*:

$$\ln N! = N \ln N - N$$

it is easy to show that [15] maximizing ln*W* is the same as maximizing entropy *H*, subject to the *N* and *M* constraints, where *H* is given by:

$$H = -\sum_{i=1}^{k} p_i \ln p_i$$

Thus, in essence, the evolution of an economic system of rational agents (*i.e.*, employees and companies) in the abstract phase space is the same as that of a thermodynamic system of molecules. Our claim is that once one transforms the dynamics of a free market process as the wanderings in the *economic* phase space, then the formal and powerful mathematical and conceptual machinery of statistical mechanics can be unleashed to prove the maximum entropy outcome at equilibrium. Thus, we conclude that the evolution of our company *A* in an ideal free market system will also be governed by the maximum entropy principle with the important distinction that entropy here is a *measure of fairness*. We discuss this interpretation and its implications in greater detail in Section 4.

Even though, for the sake of simplicity of analysis, we made some assumptions regarding the number of different employee classes, the number of employees in a class, the 50:50 compromise in negotiations, etc., these are not necessary to arrive at the conclusions we described above. It is easy to see that one could have started with a different number of classes, different number of employees, and a different kind of compromise, and would still have arrived at the same general conclusions.

*3.2. Entropy as a Measure of Fairness: Axiomatic Approach*

In the second approach, we utilize the axiomatic framework introduced by Lan *et al*. [8]. They prove that any measure of fairness has to satisfy the following properties: continuity, homogeneity, asymptotic saturation, partitional independence, and monotonicity. For proofs and other details of their framework we refer the reader their paper. Their framework is similar to the one introduced by Khinchin [14,15] in deriving entropy axiomatically as a measure of uncertainty in information theory.

Lan *et al*. introduce a set of five axioms that a fairness measure *f(S)* should satisfy:

*(i) Axiom of Continuity:* Fairness measure *f(S)* is continuous. This states that a small change in salary results in a small change in the fairness measure.

(ii) *Axiom of Homogeneity:* Fairness measure *f(S)* is a homogeneous function of degree zero. This axiom states that the fairness measure is independent of the unit of measurement or absolute magnitude of the salaries.

(iii) *Axiom of Asymptotic Saturation:* Fairness measure *f(S)* of equal resource allocations eventually becomes independent of the number of employees. This is needed to ensure uniqueness of the fairness measure and invariance under change of variable due to scaling.

(iv) *Axiom of Irrelevance of Partition:* If one partitions the elements of *S* into two parts $S = [S^1, S^2]$, then the fairness index $f(S^1, S^2)$ can be computed recursively and is independent of the partition.

(v) *Axiom of Monotonicity:* For *n = 2* employees who are contributing equally, the fairness measure $f(\alpha, 1-\alpha)$ is monotonically increasing as the absolute difference between the two elements (*i.e.*, $|1 - 2\alpha|$) shrinks to zero. This axiom involves a value statement on fairness–*i.e.*, when there are just two equal employees, more equalized is more fair. This axiom specifies an increasing direction of fairness and ensures uniqueness of *f(S)*.

From this set of axioms, and by using a generating function, Lan *et al*. derive a number of different fairness measures, including, maximin, proportional, Gini coefficient and entropy. Of these, only entropy is the appropriate measure of fairness for the ideal free market environment as it does not require the presence of a central authority to compute and enforce the fairness policy. *None* of the rational agents, be it an employee or a company, in an ideal free market system is computing any fairness measure to promote fairness. All of them are concerned only about their own selfish economic interests and making those self serving decisions, and the system thus evolves in a self-organizing manner. This rules out all other measures except for entropy as the obvious choice.

Thus, we see that both approaches lead to the same conclusion that entropy is the appropriate measure of fairness in an ideal self-organizing free market system for labor. We make no specific assumptions about supply and demand or the production functions. We only require both the supply and demand to be large enough to validate our statistical reasoning. The same is expected of the firms.

## 4. Essence of Entropy is Fairness

Those who are familiar with the income inequality literature would immediately recognize the relevance of the Thiel index here. Indeed, the Thiel index *T* is an entropy-based measure that was proposed by Henri Theil as a measure of income inequality motivated by information theoretic arguments [16]. There are a few variations of this index and a commonly used form is defined by:

$$T = \frac{1}{N}\sum_{i=1}^{N} \frac{S_i}{S_{ave}} \ln \frac{S_i}{S_{ave}}$$

where $S_i$ is the salary of *i*th employee. It is related to Shannon's entropy *H* by the following equation:

$$T = \ln N - H$$

However, the rational for its introduction has not been intuitively clear, as Conceicao and Ferreira [17] observe, since Thiel's introduction [16] of the index as:

> "{The Theil} index can be interpreted as the expected information content of the indirect message which transforms the population shares as prior probabilities into the income shares as posterior probabilities."

This has led to criticisms such as the ones voiced by Amartya Sen [6]:

> "But the fact remains that {The Theil index} is an arbitrary formula, and the average of the logarithms of the reciprocals of income shares weighted by income is not a measure that is exactly overflowing with intuitive sense."

> "Given the association of doom with entropy in the context of thermodynamics it may take a little time to get used to entropy as a good thing ('How grand, entropy is on the increase!'), ..."

Further, and most crucially, its *connection* with the self-organizing *free market dynamics* has not been made. Our central contribution in this paper is to show that the ideal free market for labor does indeed promote fairness as an emergent self-organized property and identify that entropy is the appropriate measure of this fairness. Thus, our contribution is mainly a conceptual one. We wish to bring some conceptual clarity by sorting through the different notions of fairness in economics and sociology, and identify entropy as the relevant measure of fairness in an ideal free market environment for labor. In doing so, we also wish to liberate entropy from its traditional restrictive interpretations and reveal its *essence*. The crucial insight in our contribution is the realization that the very essence of entropy is *fairness*, an insight that has not been emphasized in prior work in statistical thermodynamics, information theory, or economics.

This fairness property emerges differently in different applications. In thermodynamics, being *fair* to all accessible phase space cells at equilibrium under the given constraints (*i.e.*, assigning equal probabilities to all the allowed microstates), results in entropy as a measure of *randomness or disorder* [9,10]. This is the appropriate interpretation for this particular domain, but it obscures the essential meaning of entropy as a measure of fairness. In information theory, under the design objective of being *fair* to all messages that could potentially be transmitted in a communication channel (*i.e.*, assigning equal probabilities to them), entropy emerges as a measure of *uncertainty* [13,14]. Again, this is the right interpretation for this application, but this, too, obscures the real nature of entropy. In the design of teleological systems, being *fair* to all potential operating environments, entropy emerges as a measure of *robustness* [19]. Once again, this is the right interpretation for this domain, but this also obscures its true meaning.

In all these cases, when one is maximizing entropy one is really maximizing fairness. Thus, the common theme in all these different systems is the essence of entropy as a measure of fairness, which stems from the notion of *equality* expressed mathematically. If there are $N$ possible candidates among whom a resource is to be distributed, and if no particular candidate is to be preferred over another, then the fairest distribution of the resource is one of *equal* allocation among all of them. This quantitative mathematical relationship is at the core of the concept of fairness. Bernoulli and Laplace expressed this notion in probability theory as the *Principle of Insufficient Reason*. The generalization of this principle is the *Principle of Maximum Entropy* [20] which addresses the question: "What is the fairest assignment of probabilities of several alternatives given a set of constraints?" *Thus, the roots of entropy as a fairness measure can be traced all the way back to the Principle of Insufficient Reason.* Like the elephant in the Indian parable about six blind men and an elephant, the fairness property is seen differently in different contexts with an appropriate, but restrictive, interpretation for the problem domain. Somehow, this important insight has been missed in all these years since the discovery of entropy.

It is a historical accident that the concept of entropy was discovered in the context of thermodynamics and therefore, unfortunately, got tainted with negative notions of doom and gloom, while, ironically, it is really a measure of fairness, which is a good thing. Even its subsequent "rediscovery" by Shannon in the context of information theory did not help much, as entropy now got associated with uncertainty, again not a good thing. It is important *not to confuse* entropy as a concept from physics even though it was discovered in this context. In other words, it is not like energy or momentum, which are physics-based concepts. Entropy is really a concept of probability and statistics, an important property of distributions, whose application has been found to be useful in physics and information theory. In this regard, it is more like variance which is a property of distributions, a statistical property, with applications in a wide variety of problem areas. However, as a result of this profound, but understandable, confusion about entropy as a physical principle, one got trapped in the popular notions of entropy as randomness, disorder, doom or uncertainty, which has prevented people from seeing the deep and intimate connection between statistical theories of inanimate systems composed of non-rational entities (e.g., gas molecules in thermodynamics) and of animate, teleological, systems of rational agents seen in biology, economics, and sociology.

For instance, in the past attempts on generalizing statistical thermodynamics to economics, one typically wrote down equations in economics that mirrored and mapped expressions in thermodynamics for entropy, energy, temperature, *etc.*—but no identification or justification of entropy in terms of meaningful economic concepts was made. Just as entropy is a measure of disorder in thermodynamics and uncertainty in information theory, what does entropy mean in economics? Neither interpretation, disorder nor uncertainty, makes much sense in the economic context. Economic systems work best when they have orderly markets. Why then would anyone want to maximize disorder? Similarly, economic systems work best when there is less uncertainty. Why then would anyone want to maximize uncertainty? The inability to resolve this crucial issue has been a major conceptual hurdle for decades and thwarting real progress as seen from Amartya Sen's remarks about the Theil index.

We believe that by recognizing entropy as really a measure of fairness, which is a fundamental economic principle, and showing how it is naturally and intimately connected to the free market economic environment, our theory makes a significant conceptual advance in revealing the deep connections between statistical thermodynamics, information theory, and economics.

## 5. Maximally Fair Salary Distribution at Equilibrium

As we know from statistical thermodynamics, equilibrium is attained by a molecular system when its entropy is maximized [9,10]. Applying the same reasoning, we see that, in the case of employees, the salary distribution achieves equilibrium when its entropy is maximized with the important insight that, in this case, it is *not* randomness or uncertainty, the conventional interpretations of entropy, that is maximized but *fairness*.

So, what would this equilibrium salary distribution of maximum fairness be in an ideal free market system? This was shown recently by this author [20] to be a *lognormal* distribution given by the probability density function:

$$f(\mu, \sigma) = \frac{1}{S\sigma\sqrt{2\pi}} e^{-\frac{(\ln S - \mu)^2}{2\sigma^2}}$$

with the mean:

$$E[S] = e^{\mu + \frac{1}{2}\sigma^2}$$

and variance:

$$Var[S] = (e^{\sigma^2} - 1)e^{2\mu + \sigma^2}$$

The entropy *H* of this distribution is given by:

$$H(\mu, \sigma) = \mu + \frac{1}{2} \ln(2\pi e \sigma^2)$$

It is important to note that even though our analysis is analogous to the one in statistical thermodynamics, the resulting distribution is different from that of the ideal gas. In the case of ideal gas molecules, the corresponding energy distribution is *exponential*, whereas the wage distribution is *lognormal*. The difference is due to the differences in the objective functions and the constraints between the two systems as discussed by Venkatasubramanian [20]. Any deviation from the equilibrium wage distribution would result in less of a fair deal for the participants overall, and therefore, is not likely to happen—e.g., imagine all employees voluntarily accepting lower salaries even though they are offered higher salaries, which is unlikely to happen spontaneously. Conceptually, this is similar to molecules in an isolated system collectively retreating to and staying in a *small sub-space* of the *accessible phase space spontaneously*, which, of course, is negated by the second law of thermodynamics.

It turns out that there are two different, but related, conceptual approaches to the question of what is the maximally fair distribution of wages in an organization functioning in an ideal free market environment: (i) the *desert* perspective where one asks what salary distribution do the employees *deserve* under ideal conditions and (ii) the *free market* perspective where one asks what salary distribution does the free market dynamics *produce* under ideal conditions at equilibrium.

In this paper, we have analyzed the free market perspective while in a prior contribution [20] we explored the desert perspective. Both perspectives lead to the same result of a lognormal wage distribution at equilibrium. It is not surprising that these two results agree as the two approaches are related perspectives on the same problem. In the desert view, we approached the problem from the *information theoretic* perspective while in the free market view we did so from the *statistical thermodynamic* perspective. In engineering terminology, the former is known as the *design* perspective (a "top down" approach), while the latter as the *operational* perspective (a "bottom up" approach) of the same system. As discussed above, in both cases we are maximizing fairness under the same constraints leading, naturally, to the same outcome.

## 6. Discussion

Others, typically researchers in the econophysics community, have proposed thermodynamical models for the emergence of income and wealth distributions [21-28]. Our contribution, however, even though our methodology also utilizes concepts from statistical thermodynamics, takes an entirely different perspective by asking a very different question, namely, "Does the free market for labor care about fairness?" The fairness question has neither been raised nor answered in the past econophysics approaches.

On the other hand, there has been a great amount of work by economists on fairness but these approaches have not addressed the aforementioned question. Indeed, as we observed in the introduction, the conventional wisdom in economics is that the free market for labor cares only about efficiency and not fairness. Further, the question of "Given the various measures of fairness which one is most appropriate in a free market environment?" has, again, neither been raised nor answered before.

Thus, there is a disconnect between the econophysics and mainstream economics communities in this context. The former has proposed models inspired by statistical mechanical analogues but has not interpreted entropy in economically relevant terms - in particular, it has not addressed the issue of fairness in its theories. In contrast, the latter which has proposed many theories of fairness has not recognized the relevance of, and connected with, the statistical thermodynamic theories. Our contribution is to identify the deep connections between these two thereby integrating these apparently disparate approaches in a unified conceptual framework.

By liberating entropy from its restrictive interpretations, we have shown the deep connection between the supply-demand driven, self-organizing, dynamics of an ideal free market for labor comprising of rational utility and profit maximizing animate agents and the thermodynamics of inanimate non-rational molecules. Thus, the same conceptual framework, which we call as *statistical teleodynamics* [18,20], explains and predicts certain aspects of the behavior of both kinds of systems, rational and non-rational.

One challenging question that comes up in this context is whether the ideal free market needs to be ergodic for our theory to be applicable. Jaynes asserts [19] that his formulation of the Principle of Maximum Entropy dispenses with the requirements of ergodicity, metric transitivity, etc. However, this is not a settled issue, as observed by Sklar [29]. This question needs to be explored further for free market dynamics.

We conclude that the aforementioned analysis shows that, contrary to conventional wisdom, an ideal free market environment for labor driven by the forces of supply and demand, leads towards a more fair distribution of wages as an emergent behavior, guided by Adam Smith's invisible hand of self-organization. By exploring deep connections with statistical thermodynamics, we show that entropy is the appropriate measure of fairness in a free market environment, and is maximized at equilibrium to yield a lognormal distribution of salaries as the fairest inequality of pay under ideal conditions. This result is the economic equivalent of the Gibbs-Boltzmann distribution of the energy landscape for ideal gases and has been empirically observed for the bottom 95% of the working population.

Given the statistical nature of our theory, the lognormal prediction is not valid for small, highly entrepreneurial, organizations where a handful of employees (e.g., the founders of a start-up, sports teams, movie actors, etc.) are demonstrably much more valuable than the others. As noted, simplifying assumptions were made since our objective was to develop a general theoretical framework and identify general principles that are not restricted by domain specific details and constraints. Clearly, the next steps are to conduct more comprehensive studies of salary distributions in various organizations in order to understand in greater detail the deviations in the real world from the ideal, fairness maximizing, free market for labor. Further work is also needed to generalize this result for more realistic free market environments that have open systems, taxes, and so on.

## Acknowledgments

The author is grateful to William Masters for his valuable comments on the manuscript.

*Entropy* **2010**1530